\documentclass[twocolumn,tighten]{aastex62}

\usepackage{xcolor}
\usepackage{hyperref}
\usepackage{longtable}
\usepackage{afterpage}
\usepackage{xspace}
\usepackage{upgreek}
\usepackage{multirow}
\usepackage{amsmath}
\usepackage{graphicx}

\newcommand{\setigen}{\texttt{setigen}\xspace}

\DeclareMathOperator{\Corr}{Corr}
\graphicspath{{./}{figures/}}

\shorttitle{Detecting technosignatures using ISM Scintillation}
\shortauthors{Brzycki et al.}

\begin{document}

\title{On Detecting Interstellar Scintillation in Narrowband Radio SETI}


\newcommand{\UCB}{Department of Astronomy, University of California Berkeley, Berkeley CA 94720}
\newcommand{\CU}{Department of Astronomy and Center for Astrophysics and Planetary Science, Cornell University, Ithaca, NY 14853}

\newcommand{\SSL}{Space Sciences Laboratory, University of California, Berkeley, Berkeley CA 94720}
\newcommand{\SWIN}{Centre for Astrophysics \& Supercomputing, Swinburne University of Technology, Hawthorn, VIC 3122, Australia}
\newcommand{\GBT}{Green Bank Observatory,  West Virginia, 24944, USA}
\newcommand{\OXF}{Astronomy Department, University of Oxford, Keble Rd, Oxford, OX13RH, United Kingdom}
\newcommand{\UMAN}{Department of Physics and Astronomy, University of Manchester, UK}
\newcommand{\ATNF}{Australia Telescope National Facility, CSIRO, PO Box 76, Epping, NSW 1710, Australia}
\newcommand{\HOU}{Hellenic Open University, School of Science \& Technology, Parodos Aristotelous, Perivola Patron, Greece}
\newcommand{\BSRC}{Berkeley SETI Research Center, University of California Berkeley, Berkeley CA 94720}
 \newcommand{\USQ}{University of Southern Queensland, Toowoomba, QLD 4350, Australia}

\newcommand{\SETI}{SETI Institute, Mountain View, California}
\newcommand{\KZA}{University of Malta, Institute of Space Sciences and Astronomy, Malta}
\newcommand{\PWJD}{The Breakthrough Initiatives, NASA Research Park, Bld. 18, Moffett Field, CA, 94035, USA}

\newcommand{\newtext}[1]{\color{black} #1} 


\correspondingauthor{Bryan Brzycki}
\email{bbrzycki@berkeley.edu}

\author[0000-0002-7461-107X]{Bryan Brzycki}
\affiliation{\UCB}
\affiliation{\BSRC}

\author[0000-0003-2828-7720]{Andrew P. V. Siemion}
\affiliation{\BSRC}
\affiliation{\SETI}
\affiliation{\UMAN}
\affiliation{\KZA}

\author[0000-0002-4278-3168]{Imke de Pater}
\affiliation{\UCB}

\author[0000-0002-4049-1882]{James M. Cordes}
\affiliation{\CU}

\author[0000-0002-8604-106X]{Vishal Gajjar}
\affiliation{\SETI}
\affiliation{\BSRC}

\author[0000-0003-1515-4857]{Brian Lacki}
\affiliation{\BSRC}

\author[0000-0001-7057-4999]{Sofia Sheikh}
\affiliation{\BSRC}
\affiliation{\SETI}

\begin{abstract}

To date, the search for radio technosignatures has focused on sky location as a primary discriminant between technosignature candidates and anthropogenic radio frequency interference (RFI). In this work, we investigate the possibility of searching for technosignatures by identifying the presence and nature of intensity scintillations arising from the turbulent, ionized plasma of the interstellar medium (ISM). Past works have detailed how interstellar scattering can both enhance and diminish the detectability of narrowband radio signals. We use the NE2001 Galactic free electron density model to estimate scintillation timescales to which narrowband signal searches would be sensitive, and discuss ways in which we might practically detect strong intensity scintillations in detected signals. We further analyze the RFI environment of the Robert C. Byrd Green Bank Telescope (GBT) with the proposed methodology and comment on the feasibility of using scintillation as a filter for technosignature candidates.

\end{abstract}



\section{Introduction}
\label{sec:intro}

The Search for Extraterrestrial Intelligence (SETI) aims to answer one of the most important scientific questions: are we alone in the universe? Complementing other subfields of astrobiology in the attempt to detect life outside our planet, radio SETI strives to detect and constrain the existence of technosignatures, signals that betray the presence of intelligent extraterrestrial civilizations. 

Radio and microwave astronomy has played an important role in modern SETI since the initial suggestion by \cite{Cocconi1959} to search near the neutral hydrogen line at 1.42 GHz for continuous narrowband emission. Out of the whole electromagnetic spectrum, radio frequencies are a strong candidate for searches since such emission is expected to arise from advanced civilizations for a portion of their technological activity\footnote{Judging from the technological development of our own civilization, we expect intelligent civilizations to emit radio waves as intentional transmissions or as unintentional leakage from normal activity.}, radio photons are efficient to produce, and radio waves travel relatively unimpeded by the atmosphere, dust, and the ISM \citep{oliver1971project, siemion2014searching}. Narrowband emission is particularly tantalizing as a discriminant from natural astrophysical radio phenomena, whose emission bandwidth is usually, at minimum, hundreds of Hz at microwave frequencies due to broadening effects \citep{tarter2001search}. From the relative ease at which our own civilization produces continuous, Hz-width signals, we anticipate that extraterrestrial civilizations will similarly emit narrowband signals.

From the first dedicated radio search for technosignatures by \cite{Drake:1961}, SETI experiments have vastly expanded along multiple axes to cover larger frequency bandwidths, higher resolutions, and additional signal types \citep{werthimer1985serendip, tarter2001search, Siemion:2013, wright2014near, MacMahon:2018, Price:2018, gajjar2021breakthrough}. The Breakthrough Listen (BL) initiative began in 2016 as the most comprehensive SETI search program to date, observing with large instantaneous bandwidths at facilities across the world, including the Robert C. Byrd Green Bank Telescope (GBT) in West Virginia, USA and the CSIRO Parkes telescope in New South Wales, Australia \citep{Worden:2017, MacMahon:2018, Price:2018}.

While the technology used in radio SETI has developed and improved throughout the decades, the requirements for a theoretical technosignature detection have not changed significantly. Narrowband signals are assumed to be non-natural in origin, but there is yet an ever-present background of human-made radio interference (RFI), comprised of both ground and space-based transmissions. Having a robust way of differentiating technosignature candidates from RFI is paramount if we are to ever have a convincing detection \citep{horowitz1993five}.

The primary strategy for RFI rejection in radio SETI is sky localization. If a signal is detected in multiple telescope directions, it is considered RFI, since a bona fide extra-solar technosignature should originate from a single location on the sky. To this end, BL uses ON-OFF observations, in which different pointings on the sky are observed in a cadence according to a ABABAB or ABACAD pattern \citep{Enriquez:2017, Price:2020}. To further tighten the directional filter, we require that a signal must appear in all 3 ON (A) observations to be considered a candidate. 

For a directional filter to properly work, signals must be continuous throughout the observational cadence. Ideally, a candidate would be detected in repeat observations localized in the sky, requiring even longer signal durations. However, as in terrestrial emissions, extra-solar narrowband signals could appear pulsed and otherwise have low duty-cycles. In such cases, signals could appear in only one or two ON observations in a cadence and for a subsection of those observations, causing them to be missed by current filters. 

On the other hand, RFI can also appear in only ON observations. For example, RFI signals could exhibit intensity modulations that follow the observational cadence of 5 minutes a pointing, a false positive that would pass the directional filter. While we observe false positives like this in practice, having directional requirements still serves as an interpretable basis for determining candidates, which would induce follow-up observations for potential re-detection.

This begs the question: can we differentiate narrowband signals as RFI based on morphology alone? Since ETI signals must travel to us through interstellar space, are there effects that would be observable and sufficiently unique compared to RFI modulations?

One possibility is that radio frequency scattering effects, such as diffractive scintillation and spectra broadening, could imprint on extra-solar narrowband signals, altering them enough to be resolved and distinguished from terrestrial RFI. A signal filter based on astrophysical properties would be an important tool, when applicable, for evaluating candidate technosignatures. For signals that fail the directional filter, a scattering-based filter might preserve missed candidates; for those that pass, it would amplify the likelihood of a true detection.

Radio wave scattering has been studied extensively since the onset of radio astronomy. Weak scattering from the ionosphere and solar wind or interplanetary medium (IPM) was observed to scintillate radio emission from stars \citep{smith1950origin, hewish1964interplanetary}. Pulsars themselves were discovered during one such study, and subsequent pulsar observations revealed strong scattering from the ISM \citep{hewish1968observation, scheuer1968amplitude, roberts1982dynamic}. Since then, much of our understanding of ISM scattering has come about by observing pulsars, especially by analyzing pulse broadening and intensity fluctuations in time-frequency space \citep{narayan1992physics}. This observational work has led to models describing the stochastic nature of scintillation and broadening. 

Plasma effects on narrowband signals have been analyzed by \cite{cordes1991interstellar} and \cite{cordes1997scintillation}. Spectral broadening from the IPM has been observed in the transmissions of artificial probes and studied extensively \citep{goldstein1969superior, woo1979spacecraft, harmon1983spectral, woo2007space}. For the ISM, scintillation has been historically interesting to SETI as a factor that changes the detectability of a technosignature. Most of the time, the signal intensity is reduced, but occasionally the intensity will spike as a result of constructive interference. \cite{cordes1991interstellar} recommend multiple observations spaced in time to maximize the chance of catching at least one detection.

In this work, we investigate the parameter space of scattering relevant to narrowband radio SETI and investigate whether resolved scattering effects can be used to flag technosignature candidates in the proverbial haystack of RFI. In Section \ref{sec:theory}, we review scattering theory relevant to narrowband signals. In Section \ref{sec:detection}, we introduce methods for identifying the presence of scintillation in radio spectrogram data and for producing synthetic scintillated intensity time series. In Section \ref{sec:ne2001}, we present an approach for estimating likely scattering properties as a function of observation parameters using the NE2001 model. In addition to examining theoretical properties of scintillated narrowband signals, in Section \ref{sec:rfi}, we perform a statistical analysis on detected narrowband signals in multiple radio bands using the GBT. We compare properties of real RFI signals with those of theoretical scintillated ETI signals to determine the conditions under which scattering effects can be used as effective SETI filters. Finally, we summarize our results, discuss limitations, and give recommendations on potential scintillation-based technosignature searches in Section \ref{sec:discussion}.

While examples in this paper use certain values for observational parameters, such as observation length and time resolution, the methods developed in this work are meant to be broadly applicable to various radio observations. As such, we provide a Python library \texttt{blscint}\footnote{\url{https://github.com/bbrzycki/blscint}} that implements many of the key components of our scintillation search methodology.

\section{Scattering Theory and SETI}
\label{sec:theory}

Observational and theoretical work on radio scattering have been done to characterize both the bulk power spectrum of electron density fluctuations as well as the effect of localized ionized scattering structures along the line of sight \citep{rickett2007scintillations}. In this work, we limit our considerations to the wavenumber spectrum of ISM plasma fluctuations as a first order approximation of scattering along any line of sight. 

The dominant effect causing radio scattering in ionized plasma is refraction due to variations in electron density. The changes in refractive index give rise to changes in phase when a plane radio wave is passing through the scattering layer. These phase variations, along with path-induced phase delays, are propagated to the observer's plane, creating an interference pattern.

Since ionized plasma is a complex, stochastic medium, it is most useful to describe the power spectrum of turbulent scales. In practice, it is common to use the phase structure function:
\begin{equation}
    D_\phi(x, y) = \langle [\phi(x + x', y + y') - \phi(x, y)]^2 \rangle_{x', y'},
\end{equation}
where $x, y$ are coordinates in the scattering plane. This equation can also be expressed in terms of a vector baseline $\boldsymbol{r}=\langle x, y\rangle$, which is useful when describing interferometer measurements. For single dish measurements, this ``baseline'' is set by the relative transverse velocity $V_T$ of the diffraction pattern during an observation of length $\tau$, so that $r=V_T \tau$. Here, we assume that the pattern is effectively ``frozen,'' in that $V_T$ dominates the intrinsic random motion of material in the scattering medium. The structure function is usually taken to be a power law in wavenumber (length scale), so that 
\begin{equation} \label{eq:structurefunc}
    D_\phi(r) \propto r^\alpha
\end{equation}
for some power $\alpha$ \citep{rickett1990radio, narayan1992physics}.

The phase spectrum of the scattering medium determines the type of diffraction pattern seen by the observer, so it is important to constrain this at a high level. A common assumption is that ionized scattering media are isotropic and follow Kolmogorov turbulence, such that energy cascades from large turbulent structures with an outer length scale down to an inner length scale. Long-term pulsar observations show evidence that ISM scattering exhibits a Kolmogorov spectrum over many orders of magnitude \citep{ramachandran2006interstellar}. Kolmogorov turbulence is described by $\alpha=5/3$ in Equation \ref{eq:structurefunc}. 

Another important case of turbulence is the square-law regime, for which $\alpha=2$. This typically applies when the spatial wavenumber probed by the observation (i.e. $r=V_T \tau$) is smaller than the inner scale. This regime yields nice analytical expressions for scattering behavior, such as the spectral broadening function being a Gaussian. Some ISM scattering studies have accordingly used Gaussian models derived using $\alpha=2$ as approximations for the Kolmogorov case \citep[$\alpha=5/3$; ][]{roberts1982dynamic, cordes1986space, gupta1994refractive}. 

\subsection{Weak and Strong Scattering}

Since turbulence and scattering are inherently stochastic processes, it helps to compare characteristic scales to describe the underlying physics. 

The so-called diffractive length scale $r_\text{diff}$ is defined as the characteristic transverse distance over which the root mean square phase difference is 1 rad. This can be compared with the Fresnel radius $r_F$, which describes the size of the largest cross-section along the observer-source path for which waves arrive coherently in free space, with path-induced phase delays less than $\pi$. 

If $r_\text{diff} \gg r_F$, we are in the weak scattering regime, in which refractive phase changes are small compared to path-induced phase differences and the characteristic size of a coherent emission patch on the sky is $r_F$ \citep{narayan1992physics}. If $r_\text{diff} \ll r_F$, we are instead in the strong scattering regime, in which the characteristic coherent patch size becomes $r_\text{diff}$, and plasma-induced phase changes span many radians over the Fresnel radius. The strength of scattering depends on a variety of factors, such as the free electron number density, the strength of turbulence, the emission frequency, and the distance of the source. Along a given line of sight, the scattering strength increases and eventually transitions from weak to strong \citep{cordes1991interstellar}. The transition distance, for which $r_\text{diff} \sim r_F$, depends on the emission frequency.

In the strong scattering regime, there are two types of scintillation. Diffractive scintillation is relatively fast (on order minutes to hours) and requires a compact source, such as a pulsar, while refractive scintillation is weaker and slower (on order days to years) \citep{narayan1992physics}. Diffractive scintillation arises from multi-path propagation from emission across the scattering medium, while refractive scintillation is a larger-scale geometric effect that can itself modulate diffractive scintillation effects. Since potential narrowband ETI emission would have a compact source, we focus on strong diffractive scintillation in this paper.

The ``modulation index'' $m_d$ is the root mean square of the fractional flux variation due to scintillation. In weak scattering, $m_d \ll 1$, whereas in strong scattering, $m_d \sim 1$.

\subsection{Effects of Strong Scintillation on Narrowband Signals}
\label{subsec:scint}

Pulsar observations are effective probes of intensity scintillations in time and frequency given their persistent, broadband signals. On the other hand, since narrowband signals are by definition restricted in spectral extent, we are mostly limited to studying temporal effects. To guide the discussion, we can write a basic
model for the intensity of a scintillated narrowband signal:
\begin{equation} \label{eq:signal}
    I_\text{scint}(t) = g(t) S + N(t),
\end{equation}
where $g(t)$ is the scintillation gain, $S$ is the fixed intensity of the original signal, and $N(t)$ is the background noise.

One observable effect is that for independent observations, the detected signal intensity will follow an exponential probability density function (PDF):
\begin{equation} \label{eq:pdf}
    f_g(g) = \exp(-g) H(g),
\end{equation}
where $H$ is the Heaviside step function \citep{cordes1991interstellar, cordes1997scintillation}. If we assume a continuous-wave (CW) transmitter and think of radio waves as complex phasors, we start with signals of constant amplitude modulus. As the signal refracts at different points across the scattering medium, it picks up random phase changes. Due to multi-path propagation, many independent de-phased versions of the signal are summed together at the observing plane. The asymptotic result is that an ISM scintillated signal can be modeled as a random complex Gaussian variable, whose amplitude follows a Rayleigh distribution and whose intensity therefore follows an exponential distribution \citep{goodman1975laser}.

Another effect arising from the statistical power density spectrum of plasma turbulence is that the diffraction pattern at the observing plane has a spatial autocorrelation function (ACF) with a characteristic spatial scale $r_\text{diff}$. Though this work limits discussion to the effects on narrowband signals, strong diffractive scintillations also have a spectral ACF with a characteristic scintillation bandwidth (also known as the decorrelation bandwidth).

For a single dish telescope taking a long radio observation, the diffraction pattern will sweep across the telescope at a relative transverse velocity, so that observations display a temporal ACF in diffracted intensity. In terms of the phase structure function, the temporal ACF of $g$ is given by
\begin{equation}
    \Gamma_I(\tau) = |\Gamma_E(\tau)|^2 = \exp\left[-D_\phi(V_T \tau)\right]
\end{equation}
in the Rayleigh limit \citep{cordes1991interstellar, coles2010scattering}. Note that in this work, we use the normalized autocorrelation.

\begin{figure}
\begin{center}
  \includegraphics[width=\linewidth]{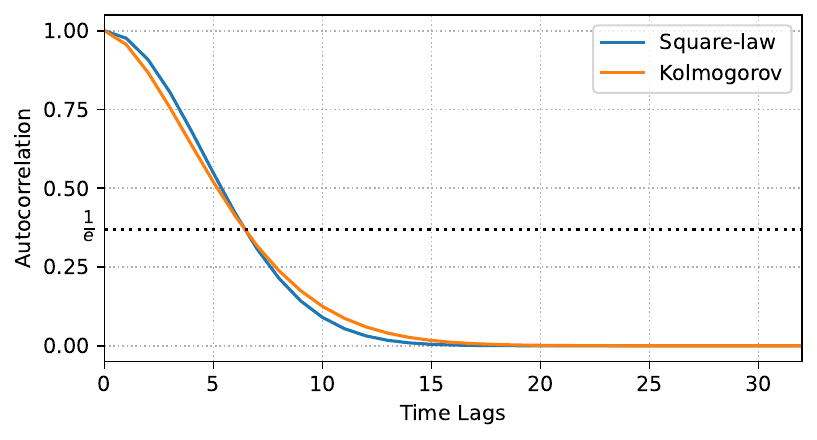}
  \caption{Comparison of the Kolmogorov and square-law ACF models. Both functions are computed using a scintillation timescale of $\Delta t_d=30\text{ s}$ and a time resolution of $\Delta t = 4.65\text{ s}$. The $1/e$-height is shown as a dotted line.}
  \label{fig:acf_models}
\end{center}
\end{figure}

The ACF thus has a representative timescale $\Delta t_d = r_\text{diff} / V_T$ over which scintillation occurs. By convention, $\Delta t_d$ is measured as the half--width at $1/e$--height of the ACF, which has been historically estimated to be a Gaussian function. In other words, 
\begin{equation} \label{eq:acfg}
    \Gamma_\text{sq}(\tau) = \exp\left[-\left(\frac{\tau}{\Delta t_d}\right)^{2}\right].
\end{equation}
However, under the Kolmogorov assumption, it is more precise to use
\begin{equation} \label{eq:acfk}
    \Gamma_\text{k}(\tau) = \exp\left[-\left|\frac{\tau}{\Delta t_d}\right|^{5/3}\right].
\end{equation}
The Kolmogorov form is near-Gaussian, as shown in Figure \ref{fig:acf_models}. In this work, we use the Kolmogorov form $\Gamma_\text{k}$ throughout, but all methods can be performed with the square-law form as well.

We note that an additional scattering effect on narrowband signals is spectral broadening. This causes power at a single frequency to spread over a bandwidth
\begin{equation}
    \Delta \nu_{sb} = C_2 / (2\pi \Delta t_d),
\end{equation}
where $C_2$ is a constant of order unity that depends on the scattering medium; $C_2 = 2.02$ is used in \cite{cordes1991interstellar}. However, at microwave frequencies, spectral broadening is typically smaller than commonly used frequency resolutions in SETI, so this effect would be difficult to observe except in lines of sight with extreme scattering. 

\section{Identifying Strong Scintillation in Detected Signals}
\label{sec:detection}

\begin{table*}
\caption{Diagnostic statistics chosen to probe theoretical scintillation effects} 
\begin{center} 
\begin{tabular}{ | c | c | c | c | } 
    \hline 
    Statistic & Data Type & Theoretical Behavior & Asymptotic Value \\ \hline \hline
    Standard Deviation (RMS) & Intensity & Exponential & 1 \\ 
    Minimum & Intensity & Exponential & 0 \\ 
    Kolmogorov-Smirnoff Statistic & Intensity & Exponential & 0 \\ 
    Autocorrelation Function $\text{ACF}(\tau)$ & Autocorrelation & Near-Gaussian & $\Gamma_I(\tau)$ \\ 
    Least Squares Fit for $\Delta t_d$ & Autocorrelation & Near-Gaussian & $\Delta t_d$\\
    \hline
  \end{tabular} 
  \tablecomments{For each statistic, we list the type of data used for computation, the theoretical behavior of that data type, and the asymptotic value of the statistic (in the absence of noise) as the observation length goes to infinity.}
  \label{table:diagstats}
\end{center}
\end{table*}

Since scintillation is inherently stochastic, we have to use statistical indicators to identify its presence in a detected narrowband signal. Accordingly, we extract time series intensity data from signals in radio Stokes I spectrograms and identify several ``diagnostic statistics'' that probe the theoretical asymptotic behavior described in Section \ref{subsec:scint}. For our scintillation analysis, we think of each signal detected within an observation of length $\tau_\text{obs}$ and spectrogram time resolution $\Delta t$ as a sequence of $N_t = \tau_\text{obs} / \Delta t$ statistically dependent random intensity samples drawn from the asymptotic distributions.

\subsection{Diagnostic Statistics}

Given time series intensity data for a detected narrowband signal, we can compute \textit{diagnostic statistics} for the expected asymptotic behavior of a scintillated signal. This process is analogous to feature engineering in machine learning, where these statistical ``features'' are designed to have a physical basis behind them. The closer a given diagnostic statistic is to the expected asymptotic value, the higher likelihood the original signal is scintillated. As such, we can create thresholds using these statistics to function as filters for interesting candidate signals.

In this paper, we offer a few examples of useful diagnostic statistics, but note that the list is in no way exhaustive and that there may be other interesting statistical features that help determine whether a given signal may be exhibiting scintillations. These can be found in Table \ref{table:diagstats}, as well as asymptotic values in the absence of noise. 

First, we want statistics that can probe the expected exponential distribution of intensities. For this discussion, assume that the time series for an idealized scintillated signal is normalized to mean 1. The standard deviation of intensity samples lends itself naturally to evaluating the degree of scintillation and tends to 1 for a normalized exponential distribution. In other words, $m_d = (\langle g(t)^2 \rangle / \langle g(t) \rangle^2 - 1)^{1/2} \sim 1$ for strong diffractive scintillation.

For a strongly scintillated signal, we expect to see complete destructive interference, leading to a minimum intensity near 0. In reality, signals are embedded in random voltage noise, so that during periods of destructive interference, measured intensities can actually be below the mean noise level. As a necessary pre-processing step to help isolate signal intensities (Section \ref{subsec:rfi_steps}), we subtract the noise mean from data spectrograms, which can result in minimum signal ``intensities'' that are negative.

Another statistical measure that addresses this directly is the Kolmogorov-Smirnoff (K-S) statistic, which is used to compare a sample distribution to a target ideal distribution using the empirical cumulative distribution function (ECDF). In this case, we compute the K-S statistic against an ideal exponential distribution with rate $\lambda=1$, keeping in mind that our time series have an assumed mean of 1. In practice, we do not know the actual mean intensities of our signals, so we can only estimate a sample mean as we normalize the time series to mean 1. So, instead of using established tables of statistic values to determine p-values, we use the statistic itself to set thresholds. The lower the K-S statistic for an intensity time series, the closer the intensities are to being exponentially distributed. 

We must note that the assumption of an unmodulated CW signal, or at least a high-duty cycle signal, is important for these statistics. For example, radio transmissions on Earth are usually modulated, so for such signals, the exponential intensity distribution arising from scintillation would be convolved with the distribution of the modulation. If the modulation is faster than the spectrogram time resolution $\Delta t$, then the modulation averages out within time bins, essentially giving us a CW signal. However, if the timescale of modulation is in between $\Delta t$ and $\tau_\text{obs}$, it is likely that the intensities of the scintillated modulated signal would no longer be exponential at the observer. 

A scintillated signal will yield a flux time series with a characteristic ACF width equal to $\Delta t_d$. From time series signal data, we can compute the ACF at all lags $k$, normalized to 1 at lag 0. We can then compare the empirical ACF with the theoretical model $\Gamma_\text{k}$ by using raw values or by fitting with least squares. In the presence of noise, the ACF spikes at lag 0 compared to non-zero lags, since the random fluctuations add in quadrature. This is especially significant for low intensity signals. Instead of only using raw (normalized) ACF values, it is therefore more reliable to fit $\Gamma_\text{k}$ and the noise spike in one shot using least squares and to derive the corresponding scintillation timescale $\Delta t_d$. Following the treatment in \citep{reardon2019modelling}, we fit the following expression to the empirical ACF:
\begin{equation}
    \Gamma_\text{k,n}(\tau) = A \Gamma_\text{k}(\tau) \Lambda(\tau, \tau_\text{obs}) + W \delta(\tau),
\end{equation}
where $A$, $W$ are multiplicative factors, $\delta$ is the Kronecker delta or discrete unit impulse function, and $\Lambda$ is the triangle function with zeros at $\pm \tau_\text{obs}$ used to model the sample autocorrelation. The least squares fit gives values for $A$, $W$, and $\Delta t_d$ within $\Gamma_\text{k}$. This process yields consistent results as if we first excluded lag 0 from the fit, which is also commonly done \citep{rickett2014interstellar}. Since detected signals may be RFI and have complex ACFs, having values for $A$ and $W$ can help us identify and exclude poor fits (i.e. if $A$ is close to 0, it is unlikely that the signal's ACF truly matches $\Gamma_\text{k}$). 

\begin{figure*}[!htb]
\begin{center}
  \includegraphics[width=\linewidth]{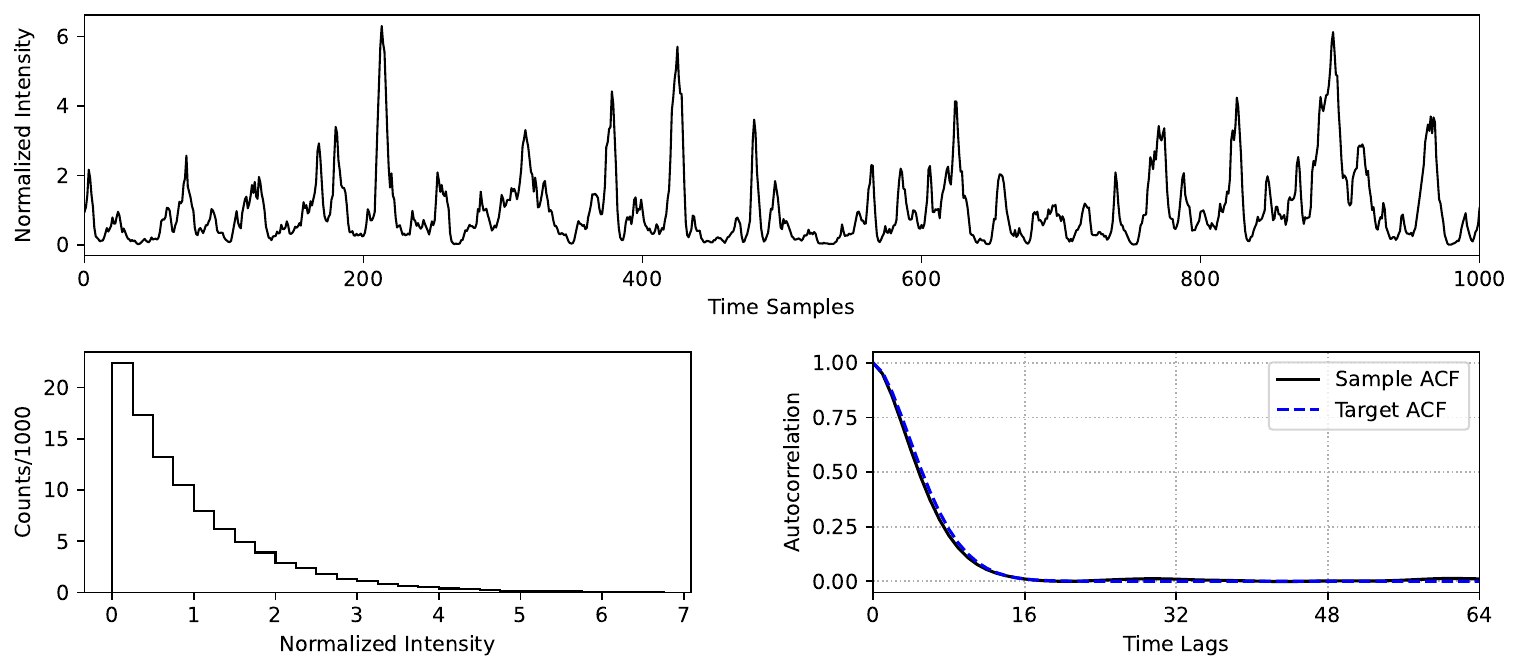}
  \caption{Synthetic scintillated intensities ($N=10^5$) generated using ARTA, using a sample interval of $\Delta t = 4.65\text{ s}$ and scintillation timescale $\Delta t_d = 30\text{ s}$. \textbf{Top}: Synthetic intensity time series data, showing first 1000 samples. \textbf{Bottom left}: Histogram of intensities, showing the expected exponential distribution. \textbf{Bottom right}: Sample ACF plotted up to lag 64, with the target ACF $\Gamma_\text{k}$ shown overlaid.}
  \label{fig:arta_sim}
\end{center}
\end{figure*}

\subsection{Constraints on Identifying Scintillation}

There are various factors at play that affect the possibility of detecting scintillation. The first is that the time resolution must be high enough to sufficiently resolve scintles (scintillation maxima). Similarly, the integration time per observation has to be long enough to collect enough scintles for better convergence to the theoretical ACF.

However, the observation length should be short enough that the receiver gain is stable. Gain fluctuations would change the underlying noise as well as the detected signal intensities over time. While this is an effect that can theoretically be corrected for using data at signal-free frequencies, for practical purposes, it is simpler to limit the observation length such that we can assume gain stability. This further avoids the potential problem of basing calculations on a ``signal-free'' region in time-frequency space that in actuality is occupied by dim RFI that escaped detection.

The detected narrowband signal must be bright enough to compute accurate statistics while embedded in noise. Noise fluctuations in the time series representation of a scintillated signal's intensity will move the empirical distribution away from exponential and mask the ACF structure. Note that since the ACF of white noise is an impulse at lag 0 and that the ACF operation is linear for uncorrelated functions, we can still fit a scaled version of the ideal profile $\Gamma_\text{k}$ for a scintillated signal's ACF, adding an additional term to fit for the noise impulse. However, for signals with low signal-to-noise ratios (S/N), the impulse will be the overwhelming part of the extracted ACF, which can make it harder to make an accurate fit. 

As one might expect in radio SETI, the RFI environment is a significant obstacle for detection. Our present tools for detecting narrowband signals make simplifying assumptions as to the kinds of signals that we hope to be sensitive to. Broadband RFI can be modulated at different frequencies, so sometimes a bright enough broadband signal passes our S/N thresholds and is falsely flagged as a ``narrowband'' detection. Broadband RFI can also overlap real narrowband signals, majorly distorting the extracted intensity time series data. It is also possible that certain modulation schemes in narrowband RFI present confounding factors for scintillation detection; perhaps some forms of RFI already appear to be scintillated (at least according to the theoretical properties identified). In Section \ref{sec:rfi}, we perform an initial analysis of the narrowband RFI environment at the GBT, computing the various diagnostic statistics and comparing them with those predicted for scintillated signals.

\subsection{Synthesizing Scintillated Signals with Autoregressive-to-Anything (ARTA)}
\label{subsec:arta}

\begin{figure*}
\begin{center}
  \includegraphics[width=\linewidth]{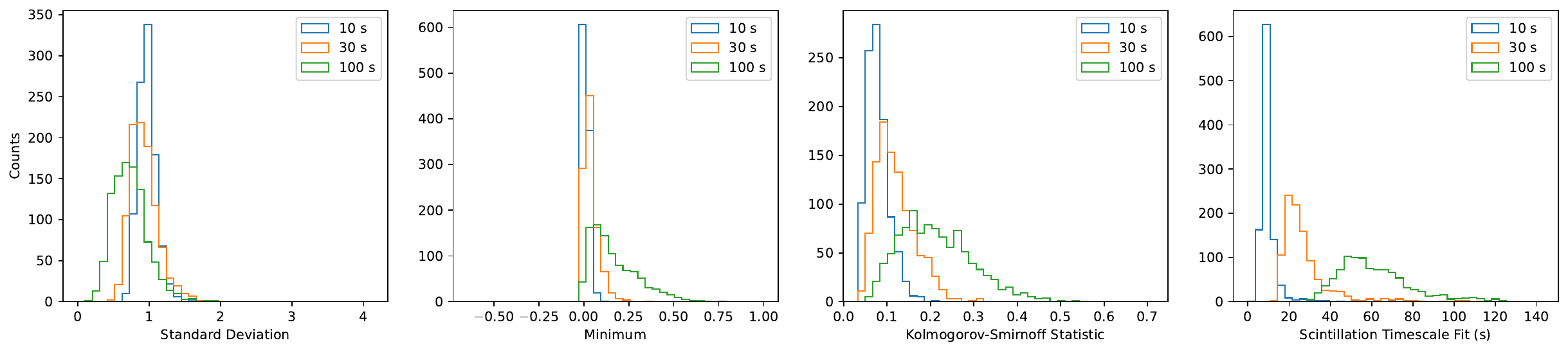}
  \caption{Histograms of diagnostic statistics computed using $N=1000$ ARTA-produced intensity time series realizations for representative scintillation timescales of 10, 30, and 100 s. Each time series is produced using $\Delta t = 4.65\text{ s}$ and $\tau_\text{obs}=600\text{ s}$ and does not include additive background noise. We plot histograms of the standard deviation, minimum, Kolmogorov-Smirnoff statistic, and least squares fit for the scintillation timescale, computed for each time series realization.}
  \label{fig:synth_stats}
\end{center}
\end{figure*}

Since observations are necessarily limited in time, we have a finite number of samples per target. Furthermore, we work with large search parameter spaces for which there is a trade-off between the length of time per target and the number of targets searched. Unless a specific pointing is otherwise scientifically interesting, it may be more useful to spend a shorter integration time on a larger number of pointings. Taken together, in most cases, we will be working with a low number of time samples per observation, which implicitly adds measurement error to each diagnostic statistic.

We would like to better understand the relationship between observation parameters, the scintillation timescale, and the expected natural error in our diagnostic statistics. Since there are a number of factors involved, it is difficult to quantify the expected errors analytically. Instead, we designed a method to create synthetic scintillated time series data, allowing us to compute the empirical distribution for each diagnostic statistic and observe the corresponding spread from the asymptotic values. 

Theoretical studies have created models of scintillation phase screens and simulated light waves passing through each screen as a function of space and frequency, such as \cite{coles1984dynamic}, \cite{hamidouche2007simulation}, \cite{coles2010scattering}, and \cite{ravi2018scintillation}. While this gives the best physical intuition for a given set of parameters, for our work, we need to be able to quickly produce a large quantity of synthetic scintillated narrowband signals over different scintillation and observation parameters. Since we are specifically interested in asking when scintillation might be detectable for SETI, we choose to rely on predictions from established theory to more efficiently create synthetic data rather than to generate our own rigorous simulations, although this may be a valuable direction for the future.

One method to produce synthetic scintillated data is to first compute the power spectrum $S$ of scintillations using a Fast Fourier Transform (FFT) of the target autocorrelation (in the voltage domain, $\Gamma_\text{k}^{1/2}$). One may then produce a complex voltage time series by taking the inverse FFT of complex Gaussian noise multiplied by $S^{1/2}$. Finally, taking the squared magnitude of the voltage series yields an intensity time series following an exponential distribution and ACF of $\Gamma_\text{k}$. While this method is relatively straightforward and satisfies asymptotic scintillation properties, we would like to present an alternative synthesis technique that may have broader uses in SETI for future applications.

Synthetic time series data following overarching statistical distributions can be produced using autoregressive models. \cite{cario1996autoregressive} developed a model called the ``autoregressive to anything'' (ARTA) process for generating time series data with arbitrary marginal distribution and autocorrelation structure (up to a specified number of lags). While this work focuses on the effects of scintillation on CW narrowband signals, having the ability to match arbitrary target distributions for first and second-order statistics could be useful for SETI applications that aim to model other astrophysical effects or even certain types of RFI. 

In our case, the target marginal distribution is exponential and the autocorrelation structure is the near-Gaussian curve $\Gamma_\text{k}$. We construct ARTA processes to model the noise-free scintillation gain $g(t)$ of a 100\% modulated narrowband signal over time. In the style of Equation \ref{eq:signal}, we can produce synthetic intensities with $I(t) = g(t) S$, for any choice of signal intensity $S$. Figure \ref{fig:arta_sim} shows an example of synthetic scintillated intensities generated in this way with $S=1$, along with a histogram and ACF plot demonstrating the asymptotic behavior.

To construct an ARTA process $Y_t$, we provide a marginal distribution with cumulative distribution function (CDF) $F_Y$ and an autocorrelation structure $\rho_Y=(\Corr[Y_t, Y_{t+1}], \dots, \Corr[Y_t, Y_{t+p}])$, where $p$ is the number of lags specified \citep{cario1996autoregressive}. Since the model is computed numerically, $\rho_Y$ is finite, and the model will only attempt to match the ACF up to lag $p$. The computation involves solving the Yule-Walker equations for a $p\times 1$ vector of autoregressive process parameters, which in turn requires inverting a $p\times p$ matrix. This limits the number of lags out to which we can effectively compute, but for scintillation analysis, this will rarely be an issue.

While this procedure results in an ARTA process with correlations close to $\rho_Y$, \cite{cario1996autoregressive} describe methods to improve convergence to the target correlations. By perturbing the input correlations to the model and doing a grid search in the parameter space, we can arrive numerically at final correlations that have higher accuracy. In this work and in \texttt{blscint} routines, we choose to forego this additional step, since it increases computational time significantly without much reward. Since using a finite observation length means that, by definition, we are performing small sample experiments, any marginal increase in the asymptotic correlation accuracy is quickly overshadowed by intrinsic sampling error. 

With this tool, for any set of parameters $(\Delta t, \tau_\text{obs}, \Delta t_d)$, we can create datasets with many time series realizations to analyze the measurement error implicit in our limited-length observations. Note that we control the observational parameters, such as $\Delta t$ and $\tau_\text{obs}$, but not the scintillation timescale $\Delta t_d$. This implies that we should choose observational parameters in such a way that we minimize our measurement error with respect to the most likely scintillation timescales. So to make this process most useful, we should attempt to estimate the most likely or most detectable scintillation timescales; this is addressed in more detail in Section \ref{sec:ne2001}.

The parameter spaces involved are vast, but we can focus on representative values close to those commonly used in radio SETI today. In other words, we try to only make slight perturbations to observational parameters used by modern spectrogram searches and similarly limit the range of scintillation timescales to practically consider. Ideally, it will be possible to directly analyze SETI observations taken for other purposes for evidence of scintillation using the methods developed in this paper.

For example, suppose we want to evaluate our sensitivity to scintillation timescales in the range of 10--100 s. The high spectral resolution data format used by BL has 2.79 Hz and 18.3 s resolution for 5 minutes, resulting in 16 time samples per observation. If we instead take observations for 10 minutes at 4.65 s resolution, yielding 128 time samples, our diagnostic statistics are more accurate and sensitive to a larger range of scintillation timescales. With these parameters, we create synthetic noise-free time series observations with ARTA, compute the diagnostic statistics, and plot histograms of each as a function of scintillation timescale as shown in Figure \ref{fig:synth_stats}.

The different scintillation timescales yield observable differences in the empirical probability density function for each diagnostic statistic. Panels 1--3 all show diagnostic statistics that target the asymptotic exponential distribution of intensities. As the scintillation timescale decreases and approaches the time resolution, each scintle will generally be covered by individual time samples. As $\Delta t_d \sim \Delta t$, the ACF structure becomes irrelevant and the observed intensity samples better match the theoretical intensity distribution. In each of Panels 1--3, the 10 s histogram is the tightest around the asymptotic statistic value, whereas the 100 s histogram has the largest spread and general deviation from the asymptotic value. As the scintillation timescale increases relative to the time resolution, more samples cover individual scintles, and so the ACF structure reduces the apparent exponentiality of the intensities within a single observation or time series realization. Panel 4 shows the least squares fit for the scintillation timescale; this similarly has the largest error for the largest scintillation timescales, since there are fewer scintles during the same observation length. Once again, note that here, the diagnostic statistics are calculated for time series intensities with no additive background noise to observe how a low sample count effects the measurement error.

\section{Exploring the Parameter Space of ISM Scintillation with NE2001}
\label{sec:ne2001}

\begin{figure}
\begin{center}
  \includegraphics[width=\linewidth]{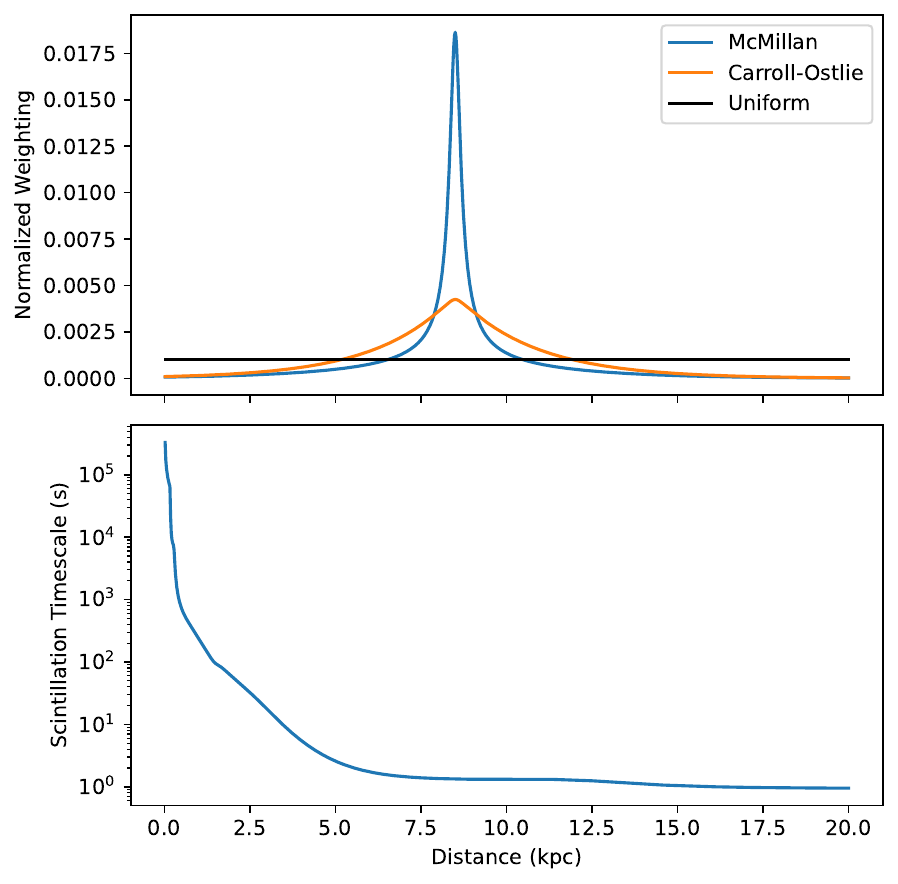}
  \caption{Comparison between methods for distance sampling, including uniformly, by stellar number density, and by stellar mass density. We use a line of sight of $(l, b)=(1, 0)$ out to a distance of 20 kpc. Bottom panel shows NE2001-produced scintillation timescales as a function of distance.}
  \label{fig:distance_weights}
\end{center}
\end{figure}

\begin{figure*}
\begin{center}
  \includegraphics[width=\linewidth]{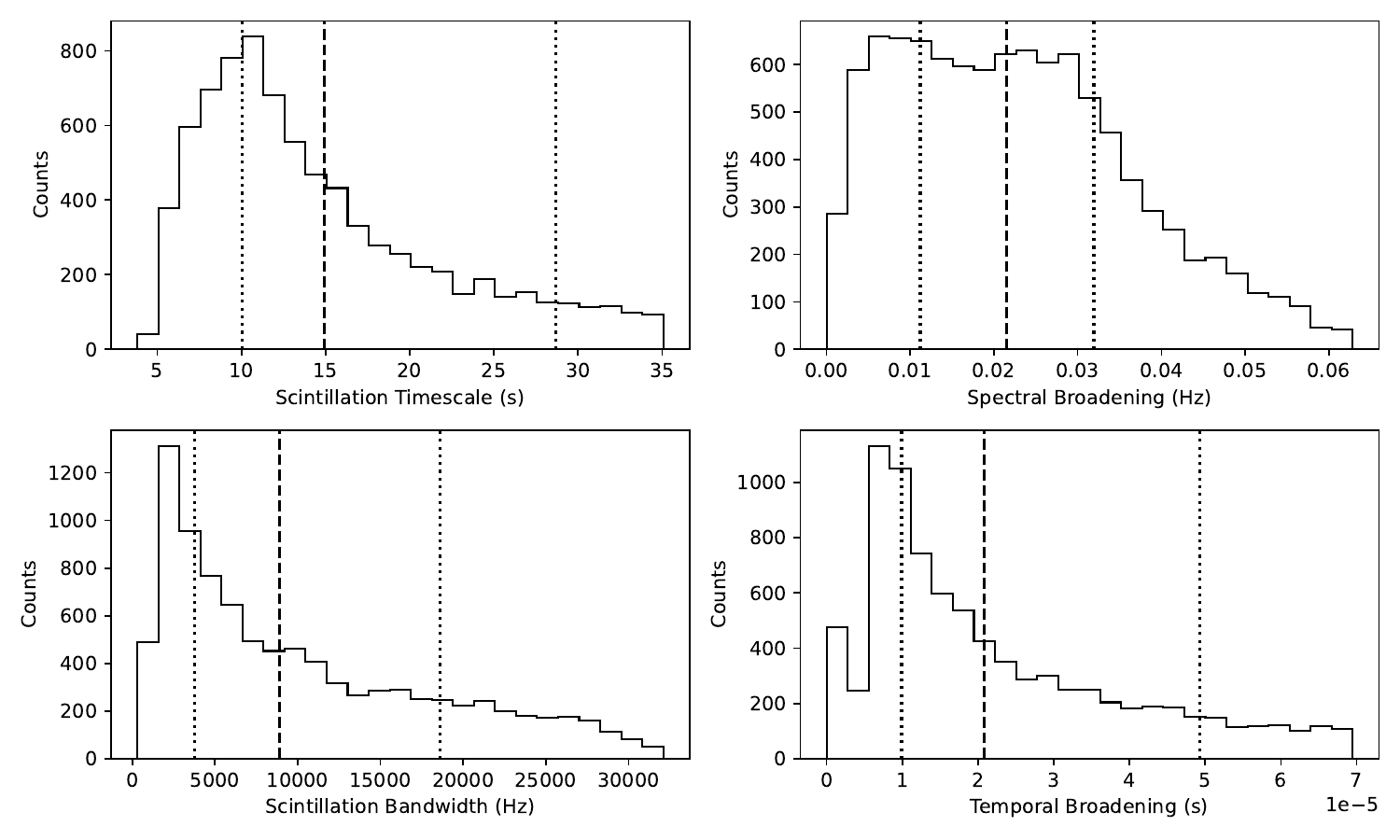}
  \caption{Set of Monte Carlo-sampled distributions of scintillation parameters at C-band, using $N=10000$ realizations. We use a line of sight of $(l, b)=(1, 0)$ out to a distance of 20 kpc, and transverse velocities are uniformly sampled between 10 to 150 km/s. Dashed line shows median value, dotted lines show interquartile range (IQR).}
  \label{fig:l1b0density}
\end{center}
\end{figure*}

The likelihood of detecting scintillation depends heavily on our physical location in our Galaxy and the lines of sight at which we observe. To determine the best targets for detecting scintillation, we need to estimate the quantitative effects of scintillation on narrowband signals in various directions on the sky. This depends on the plasma free electron number density and strength of turbulence along the line of sight. 

\cite{cordes2002ne2001} developed the NE2001 free electron density model for our Galaxy, based on pulsar observations and scattering studies. NE2001 models various Galactic features and estimates the dispersion measure (DM) and characteristic scattering scales to distance $d$ along any given line of sight through the Galaxy. The scattering scales computed include the scintillation timescale, spectral broadening, scintillation bandwidth, and temporal broadening. This allows us to uniquely estimate the asymptotic statistical properties of scintillation, which can help decide promising targets for scintillation analysis.

Given a distance $d$ and Galactic coordinates $(l, b)$, the publicly-available code for NE2001 model estimates the expected scintillation timescale and bandwidth at frequency $\nu=1\text{ GHz}$ and transverse velocity $V_T=100\text{ km/s}$. From this point, we have the scaling relation:
\begin{equation} \label{eq:scintprop}
    \Delta t_d \propto \nu^{2 / \alpha} V_T^{-1},
\end{equation}
where $\alpha=5/3$ for Kolmogorov turbulence and $\alpha=2$ for square-law turbulence \citep{cordes1997scintillation, coles2010scattering}. With Equation \ref{eq:scintprop}, we can scale raw NE2001 values to estimate scintillation properties for specific observational setups. 

We would like to narrow the parameter space of possible observing configurations and scintillation timescales to those that are most amenable to detection with current facilities. With the NE2001 model, we can estimate scintillation properties for a given set of input parameters, including the sky direction, distance, frequency, and transverse velocity. However, these inputs constitute an enormous parameter space, with no clear \textit{a priori} preference from a SETI perspective. Even with bounds for each individual parameter, it would be prohibitively computationally expensive to calculate properties across each combination of potential parameters. Instead, we choose to use Monte Carlo sampling over the parameter space, using enough samples to sufficiently capture the core statistics of the distribution of scintillation properties. 

For sampling, we fix a sky direction $(l, b)$ and a target radio frequency band. We then sample the frequency $\nu$ uniformly within that band (as a narrowband signal could be found anywhere in the band). In this paper, we will refer to common radio bands used with the GBT, including L (1.15--1.73 GHz), S (1.73--2.6 GHz), C (3.95--8.0 GHz), and X (8.0--11.6 GHz) \citep{gbt2017proposer, MacMahon:2018}.

For the distance $d$, we have to specify a maximum distance $d_\text{max}$, but the minimum distance $d_\text{tr}$ is that at which weak scattering transitions to strong scattering. We can sample uniformly from $[d_\text{tr}, d_\text{max}]$, but we can also attempt to match the potential distribution of distances that ETI would actually occur. For example, we can sample distances based on the expected distribution of stellar number densities along the line of sight through the Galaxy. For this, we use model parameters from \cite{gowanlock2011model}, who adapted a model from \cite{carroll2007introduction} that matches the observed density in the solar neighborhood. To see the effects on our sampling, we can also sample by stellar mass density, though this is less precise, since we typically expect ETI to reside around less massive stars. We use the model provided in \cite{mcmillan2016mass} to compute stellar mass density along a line of sight. In Figure \ref{fig:distance_weights}, we compare these models as a function of distance along Galactic coordinates $(l, b)=(1, 0)$, showing them alongside NE2001-generated scintillation timescales. As expected, the mass density profile is significantly sharper than the number density, but both more heavily weight the Galactic center region compared to uniform distance sampling.

Finally, the transverse velocity $V_T$ is perhaps the hardest to constrain in general. For scintillation, $V_T$ depends on the relative transverse velocities of the source, observer, and scattering screen, each of which is difficult to predict. A representative transverse velocity for Galactic pulsars is about 100 km/s \citep{cordes1986space}. The transverse velocity for an ETI source, especially in our solar neighborhood, might be on order 10 km/s instead \citep{cordes1991interstellar, cordes1998diffractive}. Depending on the line of sight, for sources far across the Galaxy (i.e. 10 kpc or so), differential Galactic rotation can add components to the transverse velocity on order of 100 km/s as well. An emitter’s orbital velocity and spin velocity can also contribute. Since all of these independent effects are non-trivial and stochastic, we can at best set heuristic transverse velocity ranges and sample uniformly between them, understanding that even the limits themselves are only useful to an order of magnitude.

Taking all these parameters together, we can create sampled distributions for each scintillation scale. Figure \ref{fig:l1b0density} shows a realization of Monte Carlo simulations for C-band in the $(1, 0)$ direction with $N=10000$ realizations, using a number density-based weighting on distance samples. We use a maximum distance of 20 kpc and a transverse velocity range of 10 to 150 km/s. It is readily apparently that the resultant distributions are significantly skewed. For example, short distances from the observer will lead to long scintillation timescales. Since the goal of the parameter space analysis is to evaluate the observational setup that gives us the best likelihood for detecting scintillation in narrowband signals, we focus on the central statistics. For skewed distributions, we choose to calculate the median and interquartile ranges (IQR) as representative values for each scale.

From Figure \ref{fig:l1b0density}, we conclude that signals at C-band in the direction $(1, 0)$ are likely to have scintillation timescales ranging between 10--28 s. Indeed, since this is the IQR, only half of the sampled timescales lie in that range, and there is an implicit bias towards the lower end of that range and below. What this really tells us is that if we are searching in that sky direction and at that frequency, we should make sure to choose observational parameters so that we are sensitive to scintillation timescales between 10--28 s. Also, note that spectral broadening is on order 0.01 Hz, which is negligible compared to typical spectral resolutions used in modern radio SETI. 

With this tool, we can estimate which range of scintillation timescales to target for a given sky direction and frequency band.

\begin{figure}
\begin{center}
  \includegraphics[width=\linewidth]{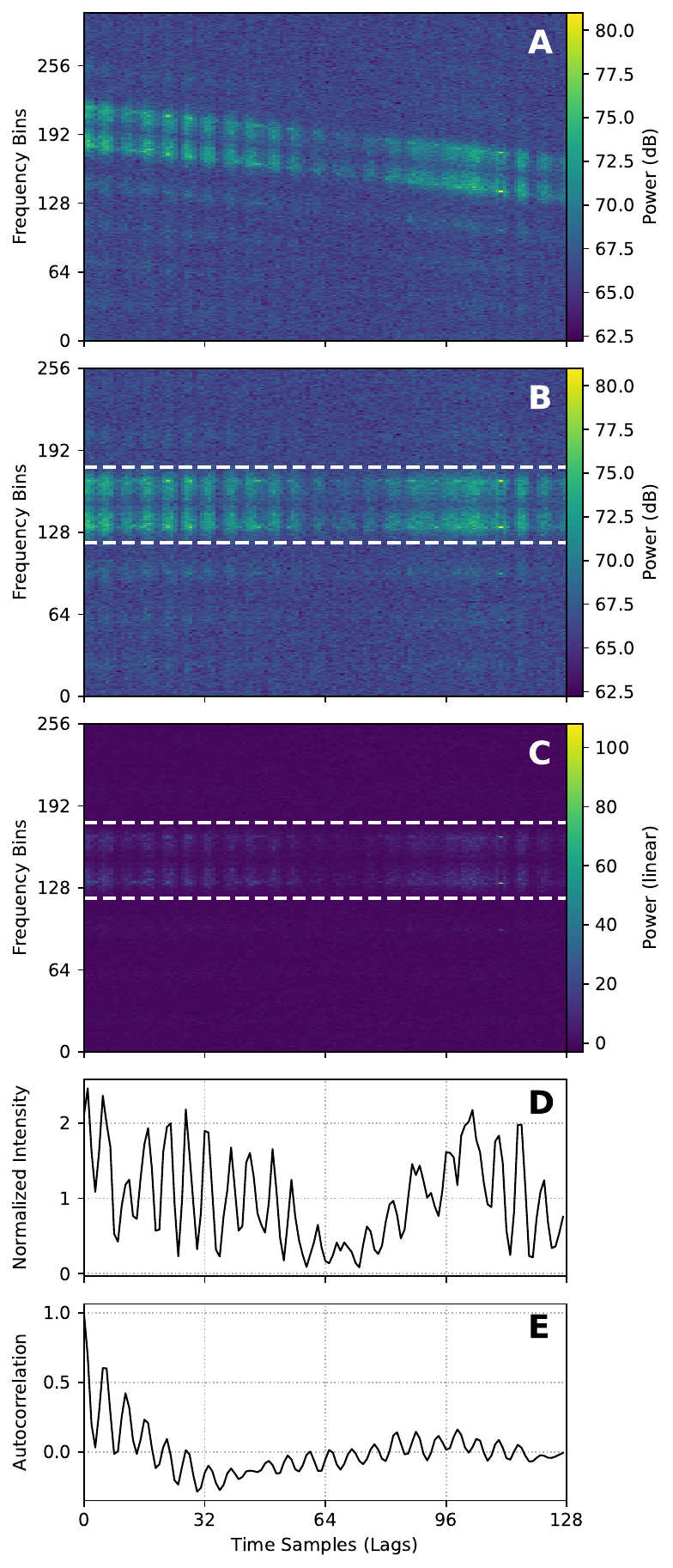}
  \caption{Steps used in signal intensity analysis. \textbf{A}: Detected narrowband signal, in GBT data. \textbf{B}: De-drifted signal from panel A, with computed bounding frequencies in dashed white lines. \textbf{C}: Frame from panel B, normalized using the background noise along the frequency axis. \textbf{D}: Time series intensities computed by integrating power in panel C between the bounding frequencies and normalized to a mean intensity of 1. \textbf{E}: Sample ACF computed from panel D.}
  \label{fig:rfi_steps}
\end{center}
\end{figure}

\section{Temporal analysis of detected narrowband RFI}
\label{sec:rfi}

To evaluate whether it is viable to detect scattering effects like scintillation in detected narrowband signals, we must characterize the standard RFI environment within which SETI observations are taken. The majority of narrowband RFI is generated from communication applications, therefore it is common for RFI to show intensity modulation in frequency or time. Depending on the nature of this modulation and the free electron column density along a line of sight, RFI could confound the detection of actual scintillated extra-solar signals. We must therefore analyze the RFI environment, regardless of sky direction, with respect to temporal statistics that can be used to identify the presence of ISM scintillation. In this paper, we focus on RFI present in GBT observations, which comprise a significant fraction of BL data.

We must note that it is technically possible that any given detected signal in this ``RFI'' analysis is actually a technosignature. However, we can confidently say that the overwhelming majority of signals encountered will be anthropogenic in origin. Furthermore, in this analysis, we take observations in a direction where $\Delta t_d$ is long compared to $\tau_\text{obs}$. This way, detected signals will not be modulated by ISM scintillation within a single observation, so whether or not a given signal is a technosignature is irrelevant to our analysis.

\subsection{Finding and Characterizing Signals}
\label{subsec:rfi_steps}

In this section, we outline the general process for detecting signals and extracting intensity time series data, from which we can compute diagnostic statistics and run our scintillation analysis. Figure \ref{fig:rfi_steps} demonstrates the step-by-step process on a real GBT RFI signal. 

The first step in analyzing the RFI environment is curating a dataset of detected signals. We need some form of energy detection to pinpoint the frequencies and preferably the drift rates of narrowband signals. The most common method for detection used by BL is the tree deDoppler code \texttt{turboSETI}\footnote{\url{https://github.com/UCBerkeleySETI/turbo_seti}}, which efficiently implements a matched filter for linearly drifting narrowband signals \citep{Enriquez:2017, enriquez2019turboseti}. \texttt{turboSETI} gives us the signal frequency at the beginning of the observation and the best-fit drift rate. However, to extract intensity data for scintillation analysis, we additionally need the frequency bandwidth that the signal occupies. 

Ultimately, we aim to construct a ``bounding box'' of sorts around each narrowband signal. Since narrowband signals can have an overarching Doppler drift rate, these bounding boxes are defined by a starting central frequency, a drift rate, and a signal bandwidth. In time-frequency space, these become bounding parallelograms, since we take the signal bandwidth to follow the extracted drift rate at each time step. Given a fit for the drift rate, we can de-drift a spectrogram containing the signal by shifting each individual spectrum accordingly, reducing the problem to finding the frequency bandwidth that overwhelmingly captures the signal's power. 

There is no singular correct way to bound radio signals found in spectrogram data. There are many morphologies of narrowband signals, such as those with unstable oscillator frequencies or varying intrinsic bandwidths. Signal leakage also affects bright signals and spreads the power into neighboring spectral bins. Background noise and nearby spurious signals can additionally complicate the bandwidth calculation. 

Signal bound estimation has been done before in radio astronomy. For pulsars, \cite{van2012pulsar} measures the size of individual pulses as the width at a user-specified fraction of the peak intensity. In one of the rare instances of bandwidth estimation in narrowband SETI, \cite{pinchuk2019search} calculates signal bounds at the $5\sigma$--level, regardless of the detected signal's peak S/N. 

Our goal is to find the tightest frequency bounds that do not exclude a significant amount of signal power, so that we can accurately represent the intensity behavior over time. If our bounds are too tight, we risk excluding and distorting information; if they are too loose, noise fluctuations can take over and wash out the signal. 

In this work, we choose to bound signals at 1\% of their maximal intensity. First, we de-drift and integrate a spectrogram along the time axis to get a spectrum centered on the signal. To make a fit of the noise background, we first exclude most of the bright data points with sigma clipping up to $3\sigma$. Then, we fit a straight line to the remaining points and obtain the final corrected spectrum by subtracting this fit from the original spectrum. The signal bounds are calculated as the frequency bins on the left and right of the signal center whose intensities dip below 1\% of the maximum intensity in the corrected spectrum. This method is balanced, capturing most of the power from signals that have apparent bandwidths ranging from a few Hz to a kHz. Figure \ref{fig:rfi_steps}B shows an example of such a fit.

To analyze the properties of a signal's intensity over time, we need to isolate the signal as best as possible from the noise background. To estimate the noise background, we use sigma clipping along the frequency axis to calculate the mean and standard deviation of noise at each timestep. We then normalize the de-drifted spectrogram at every sub-spectrum by subtracting the according noise mean and dividing by the according noise standard deviation. Theoretically, this standardizes the instrument response over the course of the observation and centers the background intensity to 0. It also serves as a crude way of filtering out simple broadband interference. Figure \ref{fig:rfi_steps}C shows the resulting spectrogram.

To get the intensity time series for a signal, we integrate the normalized spectrogram along the frequency axis between the computed frequency bounds, resulting in a 1D array of length $N_t$. To standardize the analysis, we additionally normalize this time series to have a mean of 1, as shown in Figure \ref{fig:rfi_steps}D. From the normalized time series, we compute the ACF (Figure \ref{fig:rfi_steps}E). With these two together, we can calculate all the diagnostic statistics to compare with theoretical scintillation properties. 

It is important to note that since we attempt to normalize the noise background of the spectrogram to a mean of 0 via subtraction, we may end up with negative values in our final extracted time series. Since we cannot remove the noise fluctuation entirely, the time series intensities will always be affected by noise in this way. Normalizing the time series to a mean of 1 can have the additional effect of making the negative ``intensities'' even more negative. Nevertheless, we choose to compute diagnostic statistics using the normalized time series.

\subsection{Observation Details}

\begin{figure*}
\begin{center}
  \includegraphics[width=\linewidth]{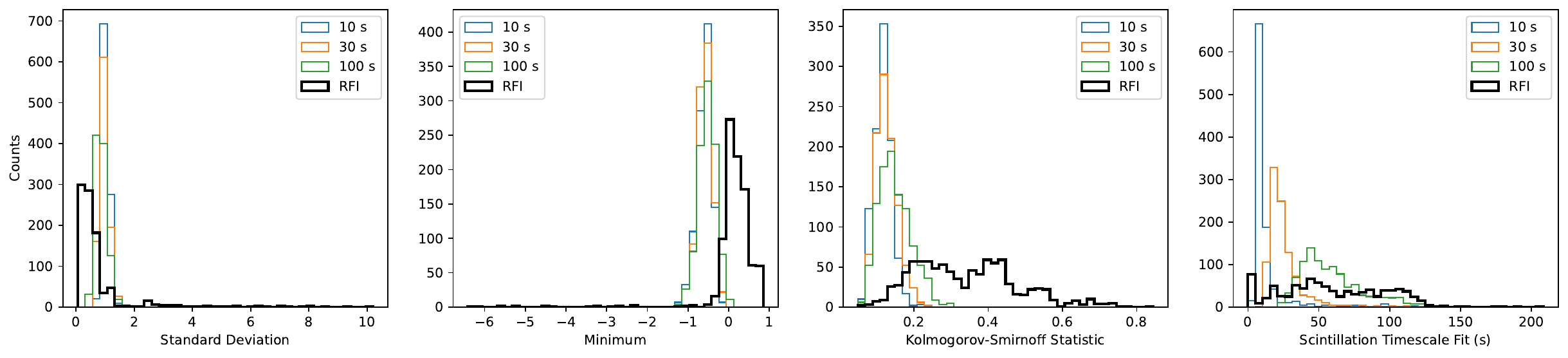}
  \caption{Histograms of diagnostic statistics for detected L-band signals with S/N$\ge$25. For each statistic, the distribution from detected RFI is shown in black. Plotted for comparison are distributions from synthetic scintillated signals at S/N=25 with scintillation timescales of 10 s (blue), 30 s (orange), and 100 s (green). Across all diagnostic statistics, it would be difficult to distinguish a true scintillated signal from RFI given the L-band RFI distributions.}
  \label{fig:L25}
\end{center}
\end{figure*}

\begin{figure*}
\begin{center}
  \includegraphics[width=\linewidth]{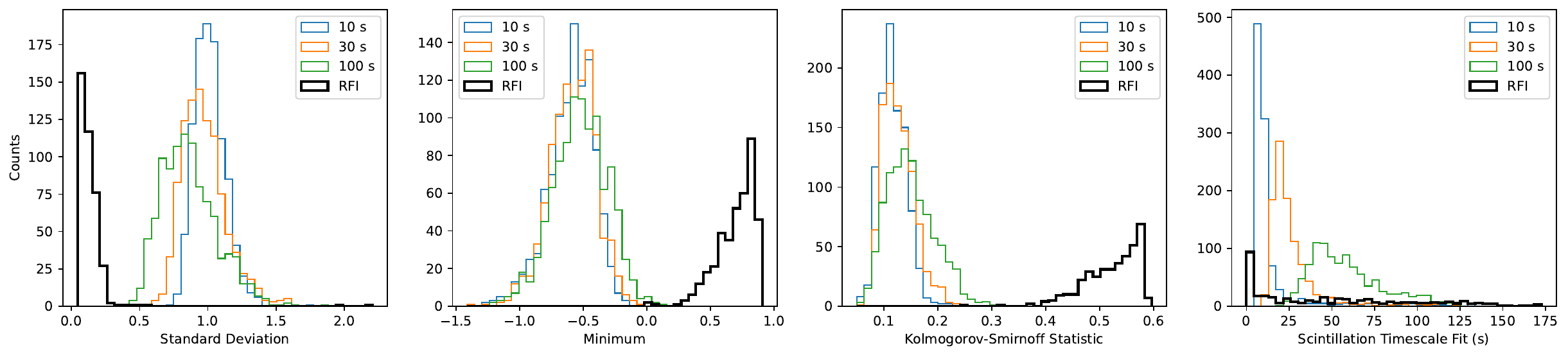}
  \caption{Histograms of diagnostic statistics for detected C-band signals with S/N$\ge$25. For each statistic, the distribution from detected RFI is shown in black. Plotted for comparison are distributions from synthetic scintillated signals at S/N=25 with scintillation timescales of 10 s (blue), 30 s (orange), and 100 s (green). It could be possible to distinguish a true scintillated signal from RFI given the C-band RFI distributions.}
  \label{fig:C25}
\end{center}
\end{figure*}

In this exploration of RFI properties, we are investigating the distribution of diagnostic statistics in real, detected RFI signals to evaluate whether these statistics can be used to identify the presence of scintillation. We must therefore ensure that our observations are unlikely to contain any scintillated signals. 

For this reason, and for additional convenience, we choose to observe towards the north celestial pole (NCP). We verified with NE2001 that the expected scintillation timescales are long compared to desired observation parameters. For instance, at 1 GHz (L-band), a signal at 1 kpc with $V_T=100$ km/s would show a scintillation timescale of 702 s. The other bands we use at the GBT (S, C, and X) correspond to even longer expected timescales due to the frequency scaling. 

The process for identifying scintillation can be performed over many observational timescales. In our case, we focus our analysis on data resolutions close to those used typically by BL. BL normally runs analysis on 5 minute integrations at a frequency resolution of 2.79 Hz and a time resolution of 18.2 s, for 16 pixels or time samples per observation. We use the same frequency resolution, but extend the data by taking 10 minute integrations at a resolution of 4.65 s, so that we get 128 samples per observation. Having more time samples leads to better diagnostic statistics and better time resolution but requires significantly more data storage. 

For this work, we used the GBT to take 10 minute observations of the NCP each at L and C-band on 2022 May 16. To find narrowband signals, we use \texttt{turboSETI} with a detection threshold of S/N=10 to search up to maximum drift rates of $\pm 5$ Hz/s. As an additional step, we exclude detections of the so-called ``DC bin'' in each coarse node, a vertical artifact of the FFT performed during fine channelization.

\subsection{Empirical Results}


Using the procedure described in Section \ref{subsec:rfi_steps}, we compute diagnostic statistics for detected signals in GBT observations taken at L and C-bands. For convenience, in this discussion, we will refer to detected GBT signals as ``RFI''. While these observations are very unlikely to contain scintillated signals, we cannot necessarily rule out the presence of technosignatures in our data. Nevertheless, we can comfortably say that the vast majority of signals are human-created interference. 

To best compare with our expectations for scintillated narrowband signals, we create synthetic GBT observations with scintillated signals produced using the methods in Section \ref{subsec:arta} and run them through the same analysis pipeline. For the synthetic signals, we construct separate datasets using $\Delta t_d$ = 10, 30, and 100 s, as in Figure \ref{fig:synth_stats}. 

The synthesis process described in Section \ref{subsec:arta} does not take noise into consideration. In this work, we treat narrowband signals as additional power that is present on top of the noise background. As such, we assume that the effects of ISM scintillation are imprinted on the signal independently from the noise background. To construct a synthetic observation, we compute a realization of a scintillated signal's intensity over time using ARTA and inject a signal with those intensities onto a radio spectrogram with a realistic noise background, following Equation \ref{eq:signal}. We use the Python package \setigen\footnote{\url{https://github.com/bbrzycki/setigen}} to inject artificial signals and compare directly with real GBT observations \citep{brzycki2022setigen}. For each scintillation timescale, we generate $N=1000$ signals with zero drift rate and the same S/N that matches our \texttt{turboSETI} detection threshold. We calculate diagnostic statistics for the artificial signals in the same way that we do for detected RFI.

The histogram comparisons for each diagnostic statistic at L and C-bands are shown in Figures \ref{fig:L25} and \ref{fig:C25}. The bold, black histograms show the non-DC RFI samples in the respective frequency band, whereas the thinner histograms represent the synthetic signal datasets. The less the RFI distributions intersect with the scintillated signal distributions, the better our methodology can distinguish a true scintillated signal. 

At a glance, C-band RFI has better separation than L-band RFI from the scintillated signal distributions, across all diagnostic statistics. In particular, for C-band, the statistics pertinent to the exponential distribution of scintillated intensities (standard deviation, minimum, K-S statistic) have relatively well-defined separations. These can be used to set thresholds (or target ranges) for each statistic, which can be combined to help filter detected signals for scintillation candidates. While the fitted scintillation timescale distributions intersect appreciably, in practice, thresholds can still be set using synthetic signal distributions and used as filters.

Comparatively, a significant portion of the L-band RFI occupies the same ranges of statistics as the synthetic signals. This means that existing RFI would confound the detection of real scintillated signals with these methods. From our observations, we observe that lower frequencies (such as L and S bands) have a relatively higher density of RFI with many morphologies, and this could be causing the distributions of statistics looking broader and more irregular than those for C-band RFI.

\section{Discussion}
\label{sec:discussion}

\subsection{Observational Recommendations for Scintillated Technosignature Searches}
\label{subsec:search}

The empirical RFI distributions suggest that at the GBT, higher frequencies will be better for creating statistics-based thresholds.\footnote{For other telescope sites, a similar RFI analysis would need to be conducted in order to draw similar insights about RFI vs. frequency.} The RFI environment at C and X-bands is less dense and less diverse than that at L and S-bands. However, scintillation effects decrease inversely with increasing frequency, lengthening the scintillation timescales (Equation \ref{eq:scintprop}). There is also a trade-off in choosing which frequencies to search: higher frequencies have more favorable RFI properties but require either longer observations or pointings with more scattering.

For each observing band, the RFI environment sets unavoidable statistics thresholds. At L-band, for instance, it is possible that there is no sky direction and no target scintillation timescale amenable for a scintillated technosignature search. While the properties of the local RFI environment certainly vary as a function of time and location, our observations suggest that lower frequencies may always be difficult to use. Specifically, the empirical L-band RFI distributions covered the ideal asymptotic value for each diagnostic statistic, implying that no variation of observational parameters could unambiguously distinguish an appreciable fraction of RFI from real scintillated signals.

On the other hand, for C-band and above, we must tend towards longer observing lengths or point towards regions of higher scattering, such as the Galactic center, in order to capture enough scintles. As discussed by \cite{gajjar2021breakthrough}, there are a multitude of reasons that an ETI detection might be most likely towards the Galactic center, making this an attractive option for a scintillated technosignature search. 

As the field of radio SETI grows and as new technosignature candidates are found, more work is being done in signal verification and follow-up analysis \citep{sheikh2021analysis, tao2022sensitive}. To this end, beyond dedicated searches for scintillation, the methods introduced in this paper may also be used as supplementary analysis for other radio SETI searches. For example, given an interesting narrowband detection that passes some SETI filters, one might ask additionally whether the signal is ISM-scintillated. Following the steps in this work and using \texttt{blscint}, one could estimate likely scintillation timescales along the observation's line of sight at the detected signal frequency. Then, one could generate synthetic ARTA datasets to set diagnostic statistic thresholds and compare how the statistics for the detected signal measure up. Assuming the signal was still compelling after these steps, it would be prudent to do a similar detected RFI analysis using the same telescope, frequency band, observation length, and time resolution to check for RFI with confounding modulation. While emission from distant sources along the Galactic plane has the best chance of exhibiting detectable scintillation within individual observations, these methods constitute a concrete framework for evaluating the likelihood of scintillation in signals from any observational radio SETI campaign.

\subsection{The Impact of Models on Designing Observational Campaigns}

The effectiveness of a designated search for scintillated technosignatures will depend on how well we can estimate the most likely values for $\Delta t_d$ as a function of sky direction and frequency. 

The fewer unknown degrees of freedom in our Monte Carlo sampling procedure (Section \ref{sec:ne2001}), the better the timescale estimates will be. For example, if we wanted to estimate what timescales are possible for emission near a particular known star, we would already begin with the location $(l, b)$ and distance $d$. The only major parameters left would be the target frequency range (which we can control) and the effective transverse velocity. By constraining sampling parameters, one can get tighter bounds for scintillation timescales and tune observation parameters accordingly. 

Our Monte Carlo procedure for scattering strength estimates relies on the NE2001 electron density model. While NE2001 remains a popular choice, the YMW16 model from \cite{yao2017new} has emerged as another prominent Galactic electron density model. There have been studies comparing both, such as \cite{deller2019microarcsecond} and \cite{price2021comparison}, particularly with regards to DM and distance estimation applied to new pulsar datasets. While YMW16 benefits from more recent data, when compared to independent pulsar measurements, both models have their own systematic estimation biases that depend on the location in the Galaxy \citep{price2021comparison}. 

The key difference for this work is that NE2001 uses scattering measurements in its fit and estimates scattering properties throughout the Galaxy \citep{cordes2002ne2001}. YMW16 specifically avoids using scattering measurements, arguing that the majority of scattering arises from relatively thin features along the line of sight and therefore cannot be used to appropriately describe the large-scale distribution of scattering \citep{yao2017new}. However, the YMW16 model still attempts to estimate pulse broadening timescales by using an empirical $\tau$--DM relation simplistically, resulting in unreliable scattering values, especially for fast radio bursts \citep{ocker2021constraining}.

While it may be difficult to develop a model that robustly constrains the effects of scattering along any line-of-sight in the Galaxy, doing so to even an order-of-magnitude would be crucial for designing scintillation search strategies for SETI, as well as for evaluating whether existing narrowband detections could benefit from scintillation analysis. As new pulsars are discovered and new Galactic electron density models are produced, we suggest that attention should still be given to scattering measurements and predictions.

\subsection{Building on the Analysis Pipeline}

While it involves many steps, the method for search and intensity extraction described in this paper is relatively straightforward. We rely on standard deDoppler search methods (e.g. \texttt{turboSETI}) to both find and characterize signal paths in one shot. Since we are searching for a stochastic effect, keeping the processing simple is not necessarily a detriment. However, our pipeline will still flag bright broadband signals that are able to exceed our S/N threshold. The philosophical question on whether a broadband impulse that contains sharp spectral features could be considered narrowband notwithstanding, using additional pre-processing to detect broadband signal features could better standardize the types of signals passing through the intensity extraction pipeline. 

Machine learning (ML) could be used to aid scintillated searches, such as for creating initial classifications of signal type and eventually even for doing final candidate analysis. In particular, deep learning techniques, such as convolutional neural networks (CNNs) have been used effectively on a variety of tasks using radio spectrograms \citep{zhang2018fast, harp2019machine, brzycki2020narrow, pinchuk2022machine, ma2023deep}. CNNs could be used to filter out spectrograms with clear broadband emission and would be relatively straightforward to integrate into the pipeline. There is certainly an avenue for complementing domain-based statistical features with computer vision methods, as is done in time-domain SETI \citep{giles2019systematic}.

ML techniques could also be applied to the extracted time series or even to the raw signal spectrogram to directly classify likely scintillation candidates. From the standpoint of interpretability, having a set of diagnostic statistics with direct links to the expected theoretical behavior of scintillated narrowband signals provides us with intuitive filter thresholds, whereas a direct ML approach might not. However, used in tandem with our methods for producing synthetic scintillated signals, supervised ML algorithms such as random forest classifiers could be used to rank each of our diagnostic statistics in their importance towards correctly distinguishing scintillated signals from RFI \citep{breiman2001random}. This could be a valuable future direction for scintillation-based searches and may very well be a function of each observatory's unique RFI environment.

\subsection{Implications and Future Directions}
\label{subsec:future}

In this work, we only focus on searching for strong scintillation on high duty-cycle narrowband signals. Since the ionosphere and IPM will tend to vary intensity relatively slightly in most cases, we identified strong scintillation from the ISM as detectable from 100\% intensity modulations. Analysis of the RFI environment at the GBT suggests that weakly scintillated extra-solar signals would be difficult to distinguish from existing interference, while strongly scintillated signals can be separated along multiple diagnostic statistics. 

A common procedure during signal verification of an interesting candidate is to search for other signals close in frequency that are similar in morphology \citep{sheikh2021analysis}. Along these lines, the possibility of simultaneous ETI signals at multiple frequencies is interesting from the perspective of a scintillation analysis. For signals separated by less than the scintillation bandwidth, we should see the same intensity modulation over time. However, for signals separated by more than the scintillation bandwidth, we would receive different intensity time series that still have the same overall scintillation timescale. With our tool to estimate scintillation timescales and bandwidths, if we were to detect multiple spectrally-nearby scintillation candidates within the same observation, we would have yet another way to contextualize the detected signals and determine whether they might actually be technosignatures.

We limit our search methodology to high duty-cycle signals, so that any fluctuations in intensity is purely due to scintillation. If an ETI transmitter is attempting to send information, the initial signal will already be modulated. This could also confound the presence of scintillation. However, we argue that along the lines of sight and distances for which we would expect narrowband signals to be scintillated, the identification of scintillation is itself a message. An ETI civilization advanced enough to transmit a message through interstellar space should understand the effects of plasma on radio emission, since it would distort the initial transmission and hinder communication. With this in mind, an ETI beacon might instead transmit a pure, unmodulated signal, expecting that other civilizations could detect the presence of scintillation in an artificial, narrowband signal. Instead of explicitly encoding a message in the narrowband signal, the mere presence of scintillation would communicate the message: ``we are here.''

Radio scattering from ionized plasma presents in other ways, such as broadband modulation and dispersion. While broadband SETI searches are relatively less common, as we explore new regions of the potential SETI signal parameter space, scintillation could be searched along the frequency axis analogously to our search along the time axis. The scintillation bandwidth, the spectral analogue of the scintillation timescale, does not vary as a function of transverse velocity, so parameter estimation may be less uncertain \citep{cordes1991interstellar}. Broadband signal searches are also able to use coarser frequency resolutions than narrowband searches, though they would likely have to use much finer time resolutions.

We hope that this work will lead to more discussion and theoretical work on other ways in which the actual radio emission that we receive can be used to identify the extra-solar origin of technosignatures. Beyond scattering, there are still properties of radio emission, such as polarization, that are only beginning to be considered in depth from a SETI perspective \citep{tao2022sensitive}. Whether it is because certain effects are stochastic or because human radio emission exploits every facet of radio light possible for communication, extracting non-trivial information from a radio signal's detailed morphology has been and will remain difficult. We may need to push the limits of detectability along hitherto unexplored axes to discover the first technosignature.

\section{Acknowledgements}

Breakthrough Listen is managed by the Breakthrough Initiatives, sponsored by the Breakthrough Prize Foundation. The Green Bank Observatory is a facility of the National Science Foundation, operated under cooperative agreement by Associated Universities, Inc. We thank the staff at the Green Bank Observatory for their operational support. S.Z.S. acknowledges that this material is based upon work supported by the National Science Foundation MPS-Ascend Postdoctoral Research Fellowship under Grant No. 2138147.


\bibliographystyle{aasjournal}
\bibliography{references}

\end{document}